\documentclass[twocolumn,showpacs,preprintnumbers,amsmath,amssymb]{revtex4}
\usepackage{graphics}
\usepackage{epsfig}
\usepackage{color}
\usepackage{subfigure}  

\begin{document}

\title{Chirped seeded free-electron lasers: self-standing light sources for two-colour pump-probe experiments}
\author{Giovanni De Ninno$^{1,2}$, Enrico Allaria$^{2}$, Benoit Mah\^{i}eu$^{1,2,3}$}
\affiliation{
1. Laboratory of Quantum Optics, Nova Gorica University, Nova Gorica, Slovenia\\
2. Sincrotrone Trieste, Trieste, Italy \\
3. Service des Photons Atomes et Mol\'{e}cules, Commissariat \`{a} l'Energie Atomique, Centre d'Etudes de Saclay, Gif-sur-Yvette, France\\
}

\date{\today}

\begin{abstract}
We demonstrate the possibility to run a single-pass free-electron laser in a new dynamical regime, which can be exploited to perform two-colour pump-probe experiments in the VUV/X-ray domain, using the free-electron laser emission both as a pump and as a probe. The studied regime is induced by triggering the free-electron laser process with a powerful laser pulse, carrying a significant and adjustable frequency chirp. As a result, the emitted light is eventually split in two sub-pulses, whose spectral and temporal separations can be independently controlled. We provide a theoretical description of this phenomenon, which is found in good agreement with experiments performed on the FERMI@Elettra free-electron laser.
\end{abstract}

\maketitle

In recent years, the advent of ultra-fast table-top laser sources has boosted the development of new experimental methodologies, based on pump-probe techniques.
In the latter, two light pulses with adjustable time delay and different wavelengths are used to investigate the processes occurring during chemical and physical reactions. The first pulse (pump) initiates the reaction, by breaking a bond or exciting one of the reactants. The second pulse (probe) is then used to interrogate the state of the reaction after a certain time delay from process start. By varying the time delay between pump and probe, and observing sample response, one is able to make a ``movie'' of the reaction. Thanks to this technique, very deep scientific and technological insight was gained in different fields, ranging from quantum communications \cite{boyer,xli}, to wave-function reconstruction in reacting molecules \cite{worn,maire}.
\\
State-of-the art table-top lasers can generate photon pulses with durations of the order of few tens of femtoseconds, thus allowing (through pump-probe methods) to gain access to the characteristic time-scale of several basic chemical and physical processes. On the other hand, the photon energies they produce is limited to the IR-visible-UV spectral range.  Such  a limitation actually prevents the possibility to reach the frontier of the length scales of the inter-molecular distances and the energy scales of the bonds holding electrons in correlated motion with near neighbours.
\\
This constraint is removed by free-electron lasers (FEL's), which are able to deliver photon pulses of femtosecond time-duration in the VUV/x-ray range \cite{macn}. However, FEL-based pump-probe techniques have to cope with practical hurdles, which pose severe limits to their development.
The strategies presently considered to set up pump-probe schemes with FEL's are basically two. In the simplest case, one can synchronize the FEL with an external laser, which ca be used as the pump, or as the probe. The first drawback of this scheme is that the possible advantage in using the short-wavelength FEL radiation is limited either to the pump or to the probe. A second issue is related to the jitter between the FEL and the external laser jitter. Being of the order of 100 fs, the latter is typically larger than achievable pulse durations. Therefore, the experimental time resolution results to be jitter-limited.  
A second possible scheme to implement pump-probe FEL experiments is based on splitting an FEL pulse, delaying the two obtained sub-pulses and recombining them at sample's location. 
This approach does not suffer from the limitations of the previous scheme. However, it does not allow to set independently the wavelengths of the pump and of the probe, since they are generated by the same FEL pulse.
\\
In this letter, we demonstrate the possibility to operate an FEL in a regime allowing to perform two-colour pump-probe experiments in the VUV/X-ray domain. In the proposed configuration, the FEL is seeded by a powerful laser pulse, carrying a significant frequency chirp. As a result, the output FEL radiation is split in two pulses, separated in time, and having different central wavelengths. Both the spectral and temporal distances between FEL pulses can be independently controlled, providing the possibility to use the FEL at the same time as a pump and as a probe. The method we propose overcomes all the issues related to the previously mentioned schemes, allowing to fully exploit the great potentiality of FEL's for pump-probe experiments.
In the following, we first introduce the basic principles on which a seeded FEL relies, and then we discuss how the FEL emission is modified in the presence of a chirped seed. Our findings are supported by a simple analytical model and by a campaign of three-dimensional numerical simulations. Analytical and numerical results are found to be in good agreement between them, and with preliminary measurements performed on the FERMI@Elettra FEL \cite{bocchetta}.
\\
\begin{figure}
\centering
{\resizebox{0.48\textwidth}{!}{ \includegraphics{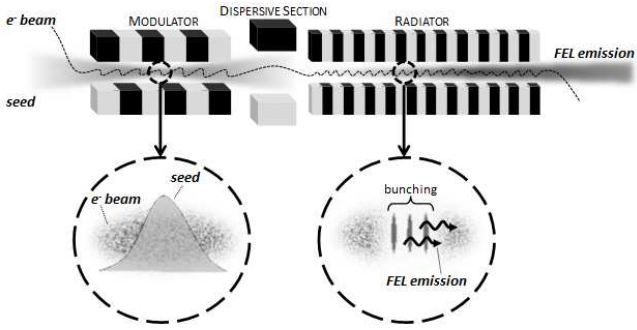}}} 
\caption{Schematic layout of a seeded single-pass FEL (harmonic generation scheme). 
\label{fig1}}
\end{figure}
Seeded FEL's can rely on different configurations \cite{yu1,yu2,alla,lambe,felda,gelo,lcls}. In this work, we focus on the so-called harmonic generation scheme, in which a relativistic electron beam propagating trough the static and periodic magnetic field generated by an undulator (called modulator) interacts with a collinear externally-injected optical pulse (seed) having wavelength $\lambda_0$, see fig.\ref{fig1}. The interaction modulates the electron-beam energy. Energy modulation is transformed into spatial bunching, when the electron beam propagates through a magnetic chicane (dispersive section). The bunching (as the energy modulation) has a periodicity equal to the seed wavelength. However, it also presents significant components at the harmonics of the latter, i.e., at $n \omega_0$ (where $\omega_0=2 \pi c/\lambda_0$, $c$ being the speed of light and $n$ an integer number). Finally, the bunched electron beam is injected into a long undulator chain (called radiator), where it emits coherently at one of the seed harmonics. In the radiator, the electromagnetic intensity generated from bunched electrons is amplified, until when, due to bunching deterioration, electrons are no longer able to supply energy to the wave and the process reaches saturation.
\\
Consider now an FEL, in which a homogeneous electron beam is seeded by a Gaussian monochromatic laser pulse, e.g., the one generated by a Ti:Sa laser. In standard operation mode, the seed peak intensity, the strength of the dispersive section and, as a consequence, the bunching at radiator entrance are tuned, so as to maximize the emission from the part of the electron beam seeded by the centre of the Gaussian pulse (see fig.\ref{fig2}a). 
The parts of the beam seeded by the tail of the Gaussian pulse will instead arrive at the radiator entrance with a local bunching smaller than optimum. As a result, the output FEL pulse will approximately reproduce the Gaussian shape of the seed, both in time and in spectrum. Suppose now to maintain constant the strength of the dispersive section, and steadily increase the seed intensity (see fig.\ref{fig2}b). For high enough intensities, the part of the electron beam which, in the modulator, interacted with the centre of the Gaussian seed will have, at the radiator entrance, a bunching larger than optimum (over-bunching). Due to the large laser-induced energy spread, the FEL emission from this part of the bunch will be significantly attenuated. The larger the seed peak intensity, the larger will be the over-bunched zone. 
The beam portions having optimum bunching (seeded by the lateral parts of the Gaussian seed) will be located both at the right and at the left of that zone. As shown in fig.\ref{fig2}b, they are disconnected.  As a consequence, in these conditions the FEL pulse will be characterized, in the time domain, by two lateral peaks or, if the seed is sufficiently intense to suppress the emission from the central part, by two separated sub-pulses \cite{labat}. Since the seed is assumed to be monochromatic, the two sub-pulses have the same wavelength. Therefore, such a regime is of little interest for (two-colour) pump-probe experiments. 
\\
As we shall see, the situation changes if the seed carries a significant frequency chirp. Indeed, in this case, the two sub-pulses will be also characterized by different wavelengths, see fig.\ref{fig2}c. This opens up the possibility of using the FEL as a self-standing source to carry out two-colour pump-probe experiments.
\\
\begin{figure}
\centering
{\resizebox{0.48\textwidth}{!}{ \includegraphics{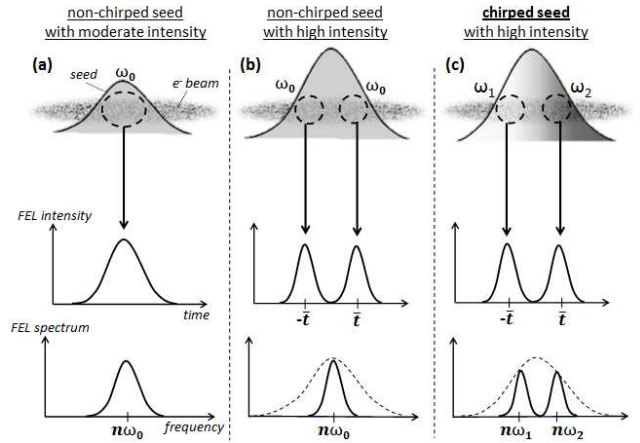}}} 
\caption{Seed-electron interaction and resulting FEL (temporal and spectral) outputs for different seed configurations: no chirp and low seed intensity (left panel), no chirp and high seed intensity (central panel), chirped seed with high intensity (right panel). The meaning of the symbol is explained in the text. 
\label{fig2}}
\end{figure}
Let's consider a seed pulse carrying a controllable quadratic frequency chirp \footnote{A controllable frequency chirp can be induced by (e.g.) propagating the seed pulse through a stretcher (prior to injection into the radiator).}. The pulse electric field reads \footnote{In eq. (\ref{campo}), the term containing the fast-varying frequency is neglected.}:

\begin{equation}
E(t) \sim \exp\left(-\Gamma_Rt^2-i\Gamma_I t^2\right)
\label{campo},
\end{equation}
where $\Gamma_R$ and $\Gamma_I$ are constants parameters controlling, respectively, the pulse duration and bandwidth. According to the previous equation, one can define the time-dependent phase of the pulse as $\phi(t)=\Gamma_It^2$.
Switching to intensity $\left(\simeq |E(t)|^2\right)$, one finds that the rms pulse temporal duration, $\sigma_l$, is given by $\sigma_l=1/\left(2\sqrt{\Gamma_R}\right)$. 
\\
The instantaneous frequency within the seed pulse, $\omega_{\mathrm inst}(t)$, is defined by the following relation:

\begin{equation}
\omega_{\mathrm inst}(t)-\omega_0=-\frac{{\mathrm d}\phi(t)}{{\mathrm d}t}=-2\Gamma_I t.
\label{omegainst}
\end{equation}

\begin{figure}
\centering
{\resizebox{0.50\textwidth}{!}{ \includegraphics{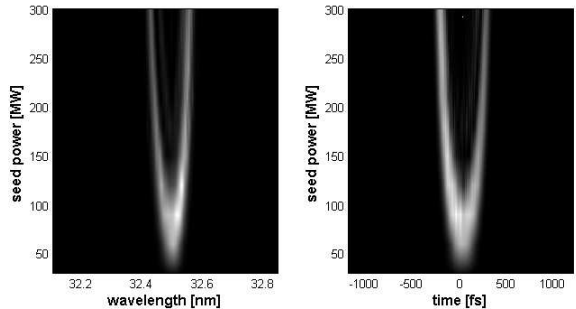}}} 
\caption{Projected spectral and temporal FEL intensities for different seed powers. For the simulation (carried out with the code PERSEO), we used the following set of parameters, corresponding to a possible configuration of the FERMI@Elettra FEL: normalized energy ($\gamma_0$)=$2.544\cdot 10^{3}$, energy spread ($\sigma_{\gamma}$)=$2.544\cdot 10^{-1}$, peak current ($I_{peak}$)= 200 A, bunch duration= 1	ps (rms); modulator period=0.1 m, modulator length ($L_u$)=3 m, modulator parameter ($K$)=5.7144, radiator period ($\lambda_w$)=$55\cdot 10^{-3}$ m, number of radiator periods ($N_u$) = 264, central modulator wavelength ($\simeq 2\pi c/\omega_0$) = 260 nm, central radiator wavelength ($\lambda_{rad}$)=32.5 nm, harmonic number ($n$)=8; seed laser power ($P_0$)=0.8-300 MW, strength of the disperive section ($R_{56}$)=54.4 $\mu$m, laser spot size in the modulator ($\sigma_r$) $\simeq 220 \mu$m, chirp ($1.08\cdot 10^{-5}\mbox{fs}^{-2}, -2.94\cdot 10^{-5}\mbox{fs}^{-2}$). 
\label{fig3}}
\end{figure}

The behaviour of a single-pass FEL seeded by a chirped laser pulse can be reliably simulated using currently available numerical codes. For the presented study, we made use of the one-dimensional code PERSEO \cite{perseo} and of the three-dimensional code GENESIS \cite{reiche}. In order to illustrate our findings, we consider the paradigmatic example of the FERMI@Elettra FEL \cite{bocchetta}, whose relevant parameters are listed in the caption of fig.\ref{fig3}. 
\\
Figure\ref{fig3} shows the spectral (left) and temporal (right) evolution of the FERMI@Elettra FEL pulse, as a function of the seed power. 
As expected, above a given power threshold the FEL pulse splits, both in time and in spectrum \footnote{It should be noted that the transition between single-pulse and double-pulse FEL behaviour is not sharp and, as a consequence the power threshold is to be defined as the power above which the (suppressed) FEL emission from the central part of electron bunch can be neglected compared to that of the two peaks. Also note that the power threshold depends on the chosen strength of the dispersive section. Generally, higher strengths correspond to lower thresholds.}. Larger seed powers correspond to larger extensions of the over-bunched zone around the pulse centre and, therefore, to larger temporal and spectral separations.  
We remark that the spectral evolution is in good agreement with preliminary experimental results \cite{private}. The maximum obtainable temporal split is limited by the electron-beam duration and/or by the possibility to generate long-enough (chirped) seed pulses characterized by significant local power at their tails.  The maximum spectral separation depends on the regime in which the FEL is operated. For relatively short longitudinal distances ($z$) inside the radiator (and/or for a relatively low electron-beam peak current), the electron beam behaves as a collection of rigid micro-bunches (see fig.\ref{fig1}), spaced by an integer number of wavelengths and radiating coherently. The field grows linearly with  $z$, so the intensity grows quadratically. This is the so-called ``low-gain'' regime. In this case, the relative FEL gain bandwidth (approximately coinciding with the relative maximum spectral separation), $B$, is proportional to the inverse of the number of radiator periods ($N_u$): $B\simeq 1/N_u$. For longer distances inside the radiator, the energy modulation of electrons induced by the FEL radiation becomes important. Hence, one can no longer consider the micro-bunches as rigid, and the power starts growing exponentially. This is the so-called ``high-gain'' regime, in which  $B\simeq \rho$, where $\rho$ is the so-called Pierce parameter \cite{boni}.
\\
Exploiting a chirped seeded FEL for carrying out pump-probe experiments relies on the possibility to control independently the spectral and temporal distance between sub-pulses. In the following we show how this can be achieved.
\\
Suppose we want to generate, at the radiator exit, two sub-pulses having constant spectral distance and variable temporal separation. First, let's see how, for a given chirp, one can decide the sub-pulses spectral distance. Call the latter $\hat{\Delta} \omega= \hat{\omega}_1-\hat{\omega}_2$, where  $\hat{\omega}_1=n\omega_1$ and $\hat{\omega}_2=n\omega_2$ are the central frequencies of the two sub-pulses, $\omega_1$ and $\omega_2$ being the corresponding instantaneous frequencies carried by the seed (determined according to eq.(\ref{omegainst})), and $n$ the selected harmonic number, see fig.\ref{fig2}.
In general, $\hat{\Delta} \omega$ will be given by $\alpha B\omega_{\mathrm rad}$, where $\alpha$ is a  factor (smaller or slightly larger than one) fixed by the user and $\omega_{\mathrm rad}=n\omega_0$ is the central frequency at which the radiator is tuned. According to eq.(\ref{omegainst}):

\begin{equation}
|\hat{\Delta} \omega|=n|\omega_2-\omega_1|=4n\Gamma_I\bar{t}\simeq \alpha n B\omega_0,
\label{deltaomega}
\end{equation}
where $\bar{t}$ is the temporal position of $\omega_1$ and $\omega_2$ (symmetric with respect to $\omega_0$, see fig.\ref{fig2}), determining the (identical) seed powers experienced by the portions of the electron beam emitting in the radiator at the two frequencies $ \hat{\omega}_1$ and $ \hat{\omega}_2$. Such a power is given by:

\begin{equation}
P=P_0\exp\left(\frac{-\bar{t}^2}{2\sigma_l^2}\right),
\label{pote}
\end{equation}
$P_0$ being the maximum seed power.
Knowing $P$, one can find the electron-beam energy modulation, $\Delta \gamma$, induced by the laser-electron interaction inside the modulator \cite{tran}:

\begin{equation}
\Delta \gamma (r)=\sqrt{\frac{P}{\bar{P}}}\frac{KL_u}{\gamma_0\sigma_r}J_{0,1}\left(\frac{K^2}{4+2K^2}\right)
\exp\left(-\frac{r^2}{4\sigma_r^2}\right).
\label{deltagamma}
\end{equation}
Here $\bar{P}\simeq 8.7$ GW, $K$ is the modulator parameter, $L_u$ the modulator length, $\gamma_0$ the nominal (normalized) beam energy, $J_{0,1}$ is the difference between the $J_{0}$ and $J_{1}$ Bessel functions, $r$  is the radial position inside the electron beam and $\sigma_r$ is the seed-laser spot size in the modulator.
The energy modulation is related to the bunching, $b$, created in the dispersive section by the following relation \cite{yu1}:

\begin{equation}
b=\exp\left(-\frac{1}{2}n^2\sigma_{\gamma}^2d\right)J_n\left(n\Delta \gamma d\right),
\label{bunch}
\end{equation}
where $\sigma_{\gamma}$  is the (normalized) electron-beam incoherent energy spread, $d$ is the strength of the dispersive section and $J_n$ the Bessel function. The parameter $d$ is usually expressed in terms of the parameter $R_{56}$, as $d=2\pi R_{56}/(\lambda_0 \gamma_0)$.
\\
Our aim is to find the optimum value of the dispersive section which, for a fixed seed power, allows maximizing the FEL emission at the two wavelengths we are interested in. Such a value cannot be found by simply maximizing the bunching parameter defined by eq. (\ref{bunch}) with respect to $d$, because in this way the emission at $ \hat{\omega}_1$ and $ \hat{\omega}_2$ would be oversaturated (and, therefore, suppressed) slightly after the entrance of the electron beam inside the radiator. Instead, one should set $R_{56}$ so as to reach maximum bunching in the last part of the radiator.
\begin{figure}
\centering
{\resizebox{0.5\textwidth}{!}{ \includegraphics{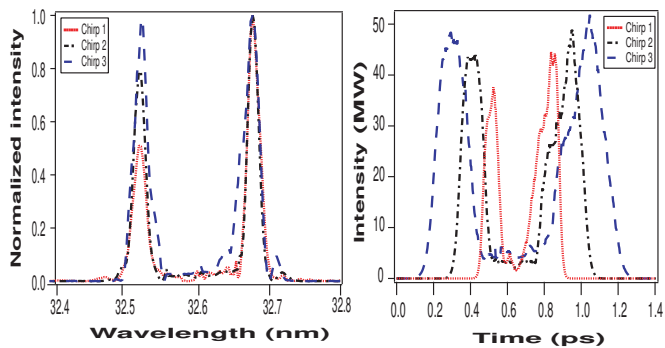}}} 
\caption{Spectral and temporal FEL output for different chirp configurations of the seed pulse: $(\Gamma_R$, $\Gamma_I)_1$=($3.18\cdot 10^{-5}\mbox{fs}^{-2}, -4.33\cdot 10^{-5}\mbox{fs}^{-2}$),$ (\Gamma_R$, $\Gamma_I)_2$=($1.08\cdot 10^{-5}\mbox{fs}^{-2}, -2.94\cdot 10^{-5}\mbox{fs}^{-2}$) and $(\Gamma_R$, $\Gamma_I)_3$=($5.13\cdot 10^{-6}\mbox{fs}^{-2}, -2.1\cdot 10^{-5}\mbox{fs}^{-2}$). For the simulation (carried out with the code GENESIS), we have used the parameters of the FERMI@Elettra FEL listed in the caption of fig.\ref{fig3}. Here $P_0=700MW$.}
\label{fig4}
\end{figure}

In order to find a criterion for determining  the optimum $R_{56}$, we assume that the FEL is operated in low-gain regime.
In this case, the additional contribution to $R_{56}$, created within the radiator at a longitudinal distance $z$, can be written as \cite{faw, alla} $2N(z)\lambda_{\mathrm rad}$, where $N(z)$ is the number of radiator periods after the longitudinal distance $z$ and $\lambda_{\mathrm rad}=2\pi c/\omega_{\mathrm rad}$ is the radiator resonant wavelength.

The value of $R_{56}$ optimising the emission at $\hat{\omega}_1$ and $ \hat{\omega}_2$, called $(R_{56})_{\mathrm opt}$, is given by
\begin{equation}
(R_{56})_{\mathrm opt}\simeq (R_{56})_{\mathrm max}-2\left(\frac{3}{4} N_u\right)\lambda_{\mathrm rad}
\label{rcsei}
\end{equation}
where $(R_{56})_{\mathrm max}$ is the value of $R_{56}$ obtained by maximising the bunching defined by eq. (\ref{bunch}). As a reasonable assumption, in the previous relation we have set $N(z)=3/4 N_u$. This in practice means that, according to eq. (\ref{rcsei}), we impose that the two portions of electron beam emitting at $\hat{\omega}_1=n\omega_1$ and $\hat{\omega}_2=n\omega_2$ reach maximum bunching after propagating through 3/4 of the radiator. The combination of the previous equations provides a simple recipe on how to use the strength of the dispersive section as a control parameter to generate, for a given chirp, two sub-pulses with fixed (decided) spectral distance. Let's now focus on the temporal domain. For constant $|\hat{\Delta} \omega|$, the temporal distance between the two peaks depends on $\bar{t}$, which is determined by the chirp parameter $\Gamma_I$ according to eq. (\ref{deltaomega}). 
\\
In conclusion, the generation of two sub-pulses with constant spectral distance and variable (adjustable) temporal separation can be be obtained by varying $\Gamma_I$, and adjusting the strength of the dispersive section as prescribed by equations (\ref{deltaomega})-(\ref{rcsei}).     
\\
As a practical example, let's consider again the case of the FERMI@Elettra FEL. Let's fix, for instance, $\hat{\Delta} \lambda(=2\pi c/\hat{\Delta} \omega) = 0.15$ nm, corresponding to $\alpha B=\hat{\Delta} \lambda/\lambda_{\mathrm rad}=4.6\cdot 10^{-3}$. Considering the three chirp configurations reported in the caption of fig.\ref{fig4}, making use of equations (\ref{deltaomega})-(\ref{rcsei}) and of the parameters reported in the caption of fig. \ref{fig3}, with  $P_0=700MW$, one finds \footnote{Note that, in the considered case, $r^2/4\sigma_r^2 \ll 1$; therefore the exponential term in eq. (\ref{deltagamma}) can be neglected.} $(R_{56})_1=30.8$ $\mu$m, $(R_{56})_2=25.6$ $\mu$m  and $(R_{56})_3=23.4$ $\mu$m. 
Carrying out GENESIS simulations using these values, one gets the spectral and temporal pulses reported in fig. \ref{fig4}.
As it can be seen, the agreement between simulations and theory can be considered quite satisfactory. Indeed, as shown in fig.\ref{fig4}a, the spectral distance between sub-pulses is quite close to 0.15 nm for all chirp values. 
As expected (see fig.\ref{fig4}b), the sub-pulses in the time domain are separated by different (adjustable) distances, depending on the chirp parameter $\Gamma_I$.         
The temporal duration depends on the local profile of the seed around the selected time $\bar{t}$: relatively flat local profiles (corresponding to $\bar{t}$ values located at seed tails) correspond to relatively long sub-pulses, while steeper local profiles (corresponding to $\bar{t}$ values closer to seed centre) will turn into shorter sub-pulses. In the absence of quadratic chirp, longer pulses should correspond to narrow bandwidths, and vice-versa. However, this trend is counteracted by the presence of the chirp on the seed. The latter is stronger in the case of longer sub-pulses and weaker when sub-pulses are shorter. As a net effect (see fig.\ref{fig4}a), the spectral bandwidth of sub-pulses remains practically constant for all chirps.   

According to simulations, the relative intensity of the two sub-pulses can be easily controlled by tuning the strength of the radiator (data not shown). Also this result has been confirmed by experiments carried out on the FERMI@Elettra FEL \cite{private}.
Simulations also show that the sub-pulses conserve memory of the seed pulse spectral chirp. This means that they could be in principle compressed in time, with a consequent significant enhancement of their peak power, without spoiling their bandwidth.

In this letter, we studied a new FEL regime, in which electrons are seeded by a powerful chirped laser pulse. As a result of this interaction, the FEL output is split in two separated pulses, both in time and in spectrum. By exploiting the interplay between seed chirp and FEL parameters, we proposed a method for controlling the temporal distance between sub-pulses, while maintaining fixed their spectral distance. For the considered parameters, the maximum temporal distance between sub-pulses is of the order of several hundreds of femtoseconds, their maximum relative spectral separation is of the order of few percent.
The analytical prediction has been successfully benchmarked with numerical simulations carried out for the case of the FERMI@Elettra FEL.  The proposed method opens up the possibility to perform two-colour pump-probe experiments in the VUV/X-ray domain using the FEL light, both as a pump and as a probe.

We gratefully acknowledge the valuable contribution of F. Benicivenga. We also profited from insightful discussions with C. Callegari, M. Coreno, M.B. Danailov, W. M. Fawley, D. Gauthier, L. Giannessi, G. Penco, K. Prince and S. Spampinati.

\end{document}